\documentclass[pre,twocolumn,floatfix,aps]{revtex4-1}
\usepackage{graphicx}
\usepackage{subfig}
\usepackage{graphicx}
\usepackage{bm}
\usepackage{amssymb}
\usepackage{amsmath}
\usepackage{color}
\begin{document}

\title{Tracer diffusion inside fibrinogen layers}

\author{Micha\l{} Cie\'sla}
 \email{michal.ciesla@uj.edu.pl}
\author{Ewa Gudowska-Nowak}
 \email{gudowska@th.if.uj.edu.pl}
\author{Francesc Sagu\'es}
 \email{f.sagues@ub.edu}
 
\author{Igor M. Sokolov}
\email{igor.sokolov@physik.hu-berlin.de}
 
\affiliation{
M. Smoluchowski Institute of Physics and M. Kac Complex Systems Research Center, Jagiellonian University, 30-059 Krako\'ow, Reymonta 4, Poland.}
\affiliation{Departament de Quimica Fisica
Universitat de Barcelona
Marti i Franques 1
E-08028 Barcelona, Spain.}
\affiliation{Institut f\"ur Physik, Humboldt Universit\"at zu Berlin, Newtonstr. 15, 12489 Berlin, Germany}
\date{\today}

\begin{abstract}

We investigate the obstructed motion of tracer (test) particles in crowded environments by carrying simulations of two-dimensional Gaussian random walk in model  fibrinogen monolayers of different orientational ordering. The fibrinogen molecules are significantly anisotropic and therefore they can form structures where orientational ordering, similar to the one observed in nematic liquid crystals, appears. 
The work focuses on the dependence between level of the orientational order (degree of environmental crowding) of fibrinogen molecules  inside a layer and non-Fickian character of the diffusion process of spherical tracer particles moving within the domain. It is shown that in general particles motion is  subdiffusive and strongly anisotropic, and its characteristic features significantly change with the orientational order parameter, concentration of fibrinogens and radius of a diffusing probe.

\end{abstract}
%
\pacs{05.45.Df, 68.43.Fg}
\maketitle

\section{Introduction}
The obstructed transport in disordered and inhomogeneous systems may lead to anomalous diffusion. Transport of a dense fluid in a porous host structure, ballistic tracer motion in a spatially heterogeneous medium, a probe  or  solute  diffusing through fixed scatterers within liquid crystalline phases  and protein motion in cellular membranes or in cytoplasm, all exhibit characteristics typical for hindered diffusion \cite{bib:Hofling,Saxton,Garini,Barkai,bib:Igor2012}.  Models of such systems usually include variants of Lorentz approach \cite{bib:Hofling} in which static scatterers are fixed in space and tracer particles are allowed to move freely through the resulting mesh. The obstacle density is crucial for the anisotropy of motion and excluded volume effects, since only unoccupied, void volume is accessible to the probe particle traveling in the system. 

In the original Lorentz model proposed as a description of the electric conductivity of metals, the motion of a ballistic particle elastically scattered off randomly distributed spherical obstacles has been analyzed. Microscopic theory predicts for this model appearance of a localization transition \cite{Goetze} where at a critical obstacle density the long-range transport ceases. In accordance, as the critical density of scatterers is approached the tracer particle experiences a sublinear in time increase of its mean-square displacement  and the diffusion constant vanishes at a critical concentration of obstacles. Such a signature of anomalous  diffusive motion is a common observation in experimental studies of various transport processes in cellular fluids \cite{Ellis}. 
Dependent on density and shapes of crowding agents, meandering subdiffusive motion occurs on larger time windows and its characteristics close to the  localization transition exhibit scaling behaviour which can be rationalized in terms of dynamic  critical phenomena \cite{bib:Hofling}.

Effects of the entrapment within the mesh composed of the obstacles and elucidation of dynamical behavior of spherical tracers diffusing through the collective structure of scatterers have been investigated so far both experimentally \cite{Holyst,Kang} and theoretically \cite{bib:Tucker,bib:Hofling,Ochab,Bagchi,Majka1,Majka2}. In particular, diffusion of silica tracer beads with large radii through colloidal rods have been observed experimentally to depend on the rod density, exhibiting domination of caging effects at high densities of rods. Similar experiments with apoferritin, quantum dots and lysozyme \cite{Holyst} diffusing in poly(ethylene glycol) solutions have been analyzed addressing a connection between the motion of nanoscopic probes
in polymer solutions and their macroviscosity. Diffusion of spherical tracers through a liquid crystal consisting of ellipsoidal particles near the isotropic-nematic phase transitions has been investigated theoretically \cite{Bagchi} pointing to the increasing anisotropy of the probe dynamics with increasing long-range order of nematogens. 
 In silico, studies performed by H\"ofling et al.\cite{bib:Hofling} analyzed dynamic scaling picture of anomalous transport in two-dimensional models, in which a Brownian tracers of different sizes were  moving between uncorrelated and overlapping circular obstacles.
These studies confirmed the universal values of the dynamic exponents of the mean-square displacement (MSD) at leading and next-to-leading order and
 the subdiffusive motion of tracers was shown to experience the crossover to normal diffusion away from the localization transition, when tracers become immobilized.

 Typically, anomalous and complex transport is conveniently described by means of MSD which, by means of the central limit theorem, is expected to grow linearly in time for time scales much larger than microscopic. On the other hand, observed non-linear growth of MSD is a benchmark of unusual, anomalous transport indicating subdiffusive behavior for 
 MSD scaling according to a power law, $\delta r^2(t)\propto t^{\alpha}$,  with an exponent $0<\alpha<1$. Apart from this
  most common characteristics of subdiffusive motion,  other phenomena like strongly reduced and time-dependent diffusion coefficients, persistent correlations in time, non-Gaussian distributions of spatial displacements and heterogeneous diffusion of particles can be also observed \cite{bib:Wang,bib:Hofling,bib:Mondiot,bib:Katja2011,bib:Igor2012,bib:Dybiec2009}.
In order to study the effects of varying geometric and spatial anisotropy in the structure of scatterers, in this article we investigate the dynamics of a single 
tracer particle performing Brownian motion in a two-dimensional mesh of randomly distributed elongated hard obstacles. At a certain critical obstacle density, the motion of the tracer becomes anomalous over many decades in time, which is rationalized in terms of an underlying percolation transition of the void space.  We analyze the scaling behavior of the time-dependent MSD and describe various facets of entrapped motion related to the tracer size and organization of the mesh.
\section{Model and simulations}
\label{sec:Model}

\subsection{Adsorption of fibrinogen molecules}
\begin{figure}
\includegraphics[width=0.6\columnwidth]{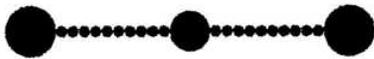}
\caption{Model of a fibrinogen molecule used in simulations: The side balls have the diameter $6.7$ nm each, the middle bead has the diameter $5.3$ nm and small spheres forming the chains are of the size $1.5$ nm. The total length of the model fibrinogen molecule is $48.7$ nm.}
\label{fig:fibrynogenmodel}
\end{figure}

Spherocylindrical objects forming scatterers are assumed to have a form of a fibrinogen molecule  (cf. Fig.\ref{fig:fibrynogenmodel})  approximated by a linear chain of stiff beads. The side beads have diameter $6.7$ nm and are connected with a central, middle-size bead of diameter $5.3$ nm  by chains of ten smaller $1.5$ nm balls .

\begin{figure}[htb]
\begin{center}
\vspace{1cm}
\includegraphics[width=0.9\columnwidth]{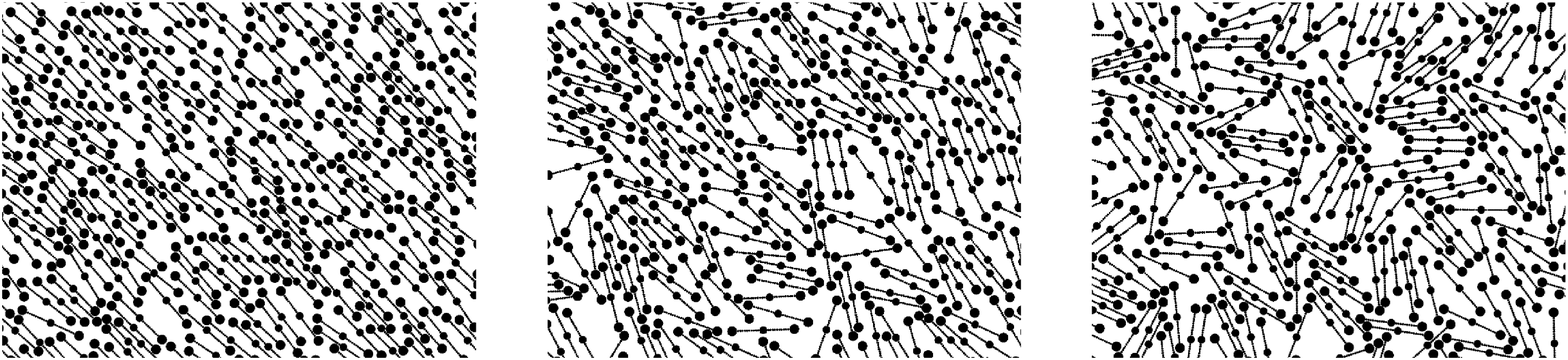}
\caption{Example coverages for $\sigma=0.1, 0.5$ and $1.0$. The mean direction $\bar{\varphi}=-\pi/4$ in all cases. The orientational order parameters derived from relation (\ref{order}) are $0.982, 0.668$ and  $q=0.188$ respectively.}
\label{fig:examples}
\end{center}
\end{figure}

Presented model had successfully explained the main properties of fibrinogen layers created during irreversible adsorption, like the saturated random coverage ratio and kinetics of the  process \cite{bib:Adamczyk2010, bib:Adamczyk2011}. The layers of different concentrations of fibrinogen are generated numerically by using the Random Sequential Adsorption (RSA) algorithm \cite{bib:Feder1980}. The procedure is based on a stochastic adsorption process in which objects are placed consecutively on a surface in such a way that they do not overlap any previously adsorbed particles and adsorb only after making a contact with an uncovered surface area. As a consequence of prohibited penetration of objects, the adsorption is completed when there is no available (free) area. The pattern of coverage of the surface obtained in such a way reflects the interplay between entropic packing of molecules and their concentration and has been displayed in Fig.\ref{fig:examples}. Accordingly, the concentration of the fibrinogen molecules $c$ depends on the orientation of the molecules, i.e. $c=c(q)$ where $q$ is the orientational parameter described below.

The single simulation run lasts for $10^4 \cdot S_C/S_F$ steps, where $S_F = 127.918$ nm$^2$ is an area covered by a single fibrinogen molecule and $S_C = 196343$ nm$^2$ is a circular surface of diameter  $500$ mn.
The saturated random coverage of the surface is determined by the dimensionless density {$NS_F/S_C=c S_F$} where $N$ represents the number of adsorbed molecules. The maximal coverage grows weakly  \cite{bib:Ciesla2012c} with increasing orientational order $q$ (the overall impact is only about 10\%) and resembles isotropic - nematic transition in a liquid crystal, where typically no significant density change is observed.

Orientational order of a fibrinogen layer is controlled here by the parameter  $\sigma$ - a width of a Gaussian probability density function, according to which the molecule's orientations are drawn. However, in order to compare results with other works and experiments it is worthy to introduce yet another order measure, independent of the mechanism that forces molecules to align in parallel. For that purpose we define a global order parameter dependent only on geometrical properties of the coverage \cite{bib:Ciesla2012a, bib:Barbasz2013a}:
\begin{equation}
q = 2\left[ \frac{1}{N}\sum_{i=1}^N \left(x_i \cos \phi + y_i \sin \phi \right)^2 - \frac{1}{2} \right].
\label{order}
\end{equation}
Here $N$ is a number of molecules in a layer, $[x_i, y_i]$ describes a unit vector parallel to the orientation of $i$-th fibrinogen and $\phi$ denotes the order's direction. Parameter $q$ is $0$ for totally disordered phase and rises with growing orientational order up to $1$. In consequence, the maximum value of $q$ is reached when all molecules are parallel.
\par
The main properties of ordered fibrinogen coverages were discussed in \cite{bib:Ciesla2012c}. Here we focus on diffusion processes in such layers. Therefore, an appropriate model of random walk process has to be introduced. 
\subsection{Diffusive transport}

Diffusion inside fibrinogen layer has been studied using Gaussian Random Walk (GRW) process. The process starts at a randomly chosen, uncovered point.  Single step is given by a random vector $ \vec{v} = [x, y]$ with coordinates drawn (as two independent random variables) according to a normal Gaussian ${\cal{N}}\left( 0, (0.35 \mbox{ nm})^2 \right)$ probability distribution. The step is accepted if it follows to an uncovered place. Otherwise, it is treated as the $[0, 0]$ step. The most typical property of such a random walk process is the mean square displacement (MSD) of the walker $\left< x^2 \right> = \left< \left| \sum \vec{v_i} \right|^2 \right>$ and its characteristic exponent $\alpha$ which reflects the dependence of the asymptotic diffusion process on time $t$:
\begin{equation}
\langle x^2 \rangle = D t^{\alpha} \Leftrightarrow \ln \langle x^2 \rangle = \ln D  + \alpha \ln t,
\label{diffusion}
\end{equation}
Here $D$ is a diffusion coefficient and $\alpha$ describes the character of the motion: for a standard (normal) diffusion MSD is expected to grow linearly in time, i.e. $\alpha=1$. 
In contrast, non-linear growth of the MSD is usually taken as an indicator of anomalous diffusive transport with slower than for the normal diffusion increase of MSD in time
($\alpha <1$) typical for the subdiffusion and the enhanced superballistic (superdiffusive) transport corresponding to  $\alpha> 1$.  This classification requires however some additional care, when diffusive motion  is defined as the asymptotic property of random walks, as e.g. charge transport in disordered semiconductors or diffusion in chemical space for random walks over polymer chains. In such realms, the combination of long steps and long breaks between them can be responsible for scaling properties of moments which are in line with ordinary Brownian motion MSD ($\alpha=1$) despite the asymptotic process is significantly non-Markov and different from Gaussian \cite{KlaSo,bib:Dybiec2009,bib:Katja2011}.
\par
Characteristics of diffusive motion, namely coefficients $D$ and $\alpha$, can be obtained directly from numerical simulations by fitting the power law (\ref{diffusion}) to measured dependence of $\langle x^2 \rangle$ on $t$. Clearly, in such case the coefficient $D$ carries different units for different exponents $\alpha$s, $[D]=[m^2/s^{\alpha}]$. Otherwise, the time-scale dependent diffusion coefficient can be introduced
\begin{equation}
D(t) = \frac{d \langle x^2 \rangle}{dt}
\label{altD}
\end{equation}
and, by analogy, the apparent exponent $\alpha$  can be defined as a logarithmic derivative of the MSD:
\begin{equation}
\alpha(t) = \frac{d \ln \left( \langle x^2 \rangle/L^2 \right) }{d \ln \left(t/T\right)}.
\label{altA}
\end{equation}
Here $L$ and $T$ are arbitrary units of length and time, respectively. Note that $\alpha$ obtained from  (\ref{diffusion}) is the mean of above $\alpha(t)$ over a finite time window used for fitting purposes. In our simulations
 $T$ represents a single step of the GRW trajectory with a typical jump length $L=0.35 nm$ (see below).
\subsection{Different diffusion models}
\begin{figure}[htb]
\vspace{1cm}
\includegraphics[width=0.4\columnwidth]{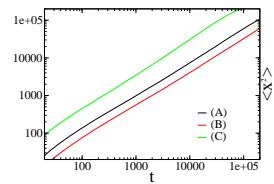}
\caption{Mean square displacement $\left< x^2 \right>$ dependence on time for three different models of a tracer diffusion: (A) - the original model, (B) -  a constant step size model, (C) - a "twice standard deviation" model. The power law fits for MSD in the  log-log scale  are:   (A)  - $\langle x^2 \rangle = 1.847\cdot t^{0.901}$, (B) - $\langle x^2 \rangle = 1.062\cdot t^{0.897}$ and (C) -$\langle x^2 \rangle = 5.1782 \cdot x^{0.942}$. The tracer diameter has been set to $0.75$ nm and the order parameter $q=0.98$.}
\label{fig:3models}
\end{figure}

In order to check whether obtained results depend on a particular simulation model of the diffusion process, two additional models of diffusive transport have been analyzed. In the first one the step size of a tracer is not drawn from the Gaussian PDF but instead, is preset to a constant $\sqrt{\delta x^2+\delta y^2} =0.35$ nm with the director of the motion sampled from a uniform distribution. The second model, similarly to case (A) uses the  Gaussian PDF of steps, however with twice bigger standard deviation (instead of $\sigma=0.35$ nm,  the value  $\sigma=0.7$ nm has been used). The total MSD $\left< x^2 \right>$ and its dependence on time for all  three models are displayed in Fig.\ref{fig:3models}.
Derived results indicate that subdiffusion inside ordered fibrinogen layers  ($q=0.98$) is not essentially model-dependent. Although diffusion coefficient $D$ differs, the slopes of (A) and (B) lines corresponding to subdiffusive $\alpha$ exponents do not differ significantly for both models. Slightly larger values of $D$ and $\alpha$  for (C) line, when a step size of the GRW was (on average) twice larger than in cases (A) and (B) can be explained by nonzero probability of jumping over the obstacles, which effectively speeds up the diffusion process. 

\section{Results}

\subsection{Short time limit}
\label{short}

\begin{figure}[htb]
\vspace{1cm}
\includegraphics[width=0.4\columnwidth]{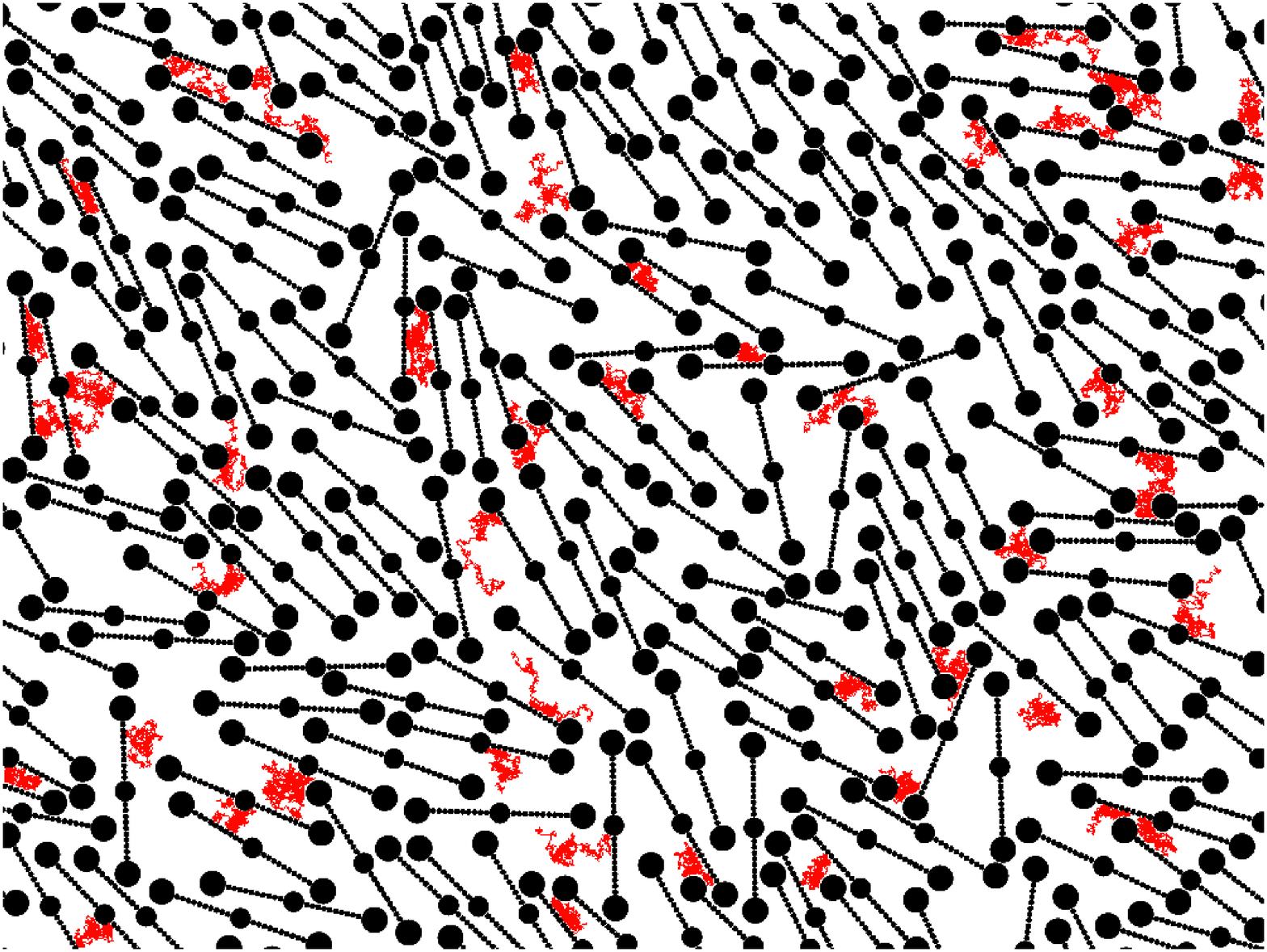}
\caption{Exemplary diffusion trajectories of 50 independent tracers inside a moderately ordered layer. Length of the paths are limited to $10^3$ steps. The orientational order measured for this layer is $q=0.67$.}
\label{fig:example_diffusion}
\end{figure}

For GRW trajectories of length equal to $10^3$ subsequent steps, the MSD has been estimated  as an average over $300$ independent realizations of the walk starting from different, randomly chosen points distributed in the void space of surface with saturated random coverage of fibrinogen molecules  (cf. Section II A). We have observed that derived values of $D$ and $\alpha$ differ significantly for various fibrinogen layers generated for the same value of $\sigma$. Therefore, those parameters have been further averaged over ensembles of $50$ up to $100$ different fibrinogen layers. Fig.\ref{fig:example_diffusion} displays exemplary diffusive trajectories of small, spherical tracer molecules wandering inside the accessible space of the fibrinogen mesh and starting their motion from randomly chosen location.

\begin{figure}[htb]
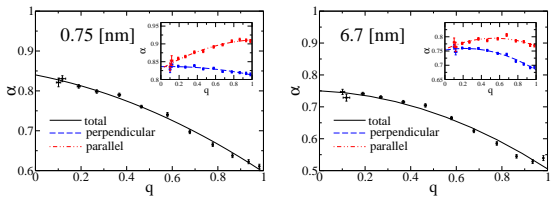

\vspace{1cm}
\centerline{
\includegraphics[width=0.4\columnwidth]{order_2}
\hspace{0.1cm}
\includegraphics[width=0.4\columnwidth]{order_9}
}
\caption{Dependence of the exponent $\alpha$ on orientational order parameter $q$ for diffusive motion of tracers of different dimension. Left (right) panel: a tracer is modeled as a spherical molecule of diameter  $0.75$ nm and $6.7$ nm, accordingly. Analysis has been performed on trajectories of $10^3$ steps of "time" duration. Insets present index $\alpha$  analyzed separately for parallel and perpendicular components of the diffusive motion. Dots represent simulation data, lines have been fitted to guide the eye.}
\label{fig:diffusion1}
\end{figure}

To relate character of the diffusion process inside fibrinogen layers to the degree of order, we have first concentrated on evaluation of the effective exponent $\alpha$ in relation (\ref{diffusion}). It's dependence on orientational order parameter $q$ is displayed in Fig.\ref{fig:diffusion1}.

\begin{figure}[htb]
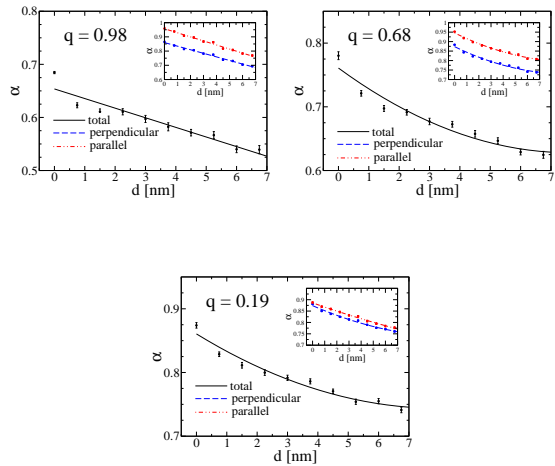

\vspace{1cm}
\centerline{
\includegraphics[width=0.4\columnwidth]{size_01}
\hspace{0.1cm}
\includegraphics[width=0.4\columnwidth]{size_05}}
\vspace{1cm}
\centerline{
\includegraphics[width=0.4\columnwidth]{size_10}}
\caption{Dependence of the effective exponent $\alpha$  on size of the tracer analyzed  for three different orientational order parameters: $q=0.98$, $0.68$ and $0.19$. Analysis has been performed on trajectories of $10^3$ steps of "time" duration. Insets display $\alpha$ characterizing parallel and perpendicular diffusion in function of the mean global orientational order direction. Dots represent simulation data, full lines have been drawn to guide the eye.}
\label{fig:diffusion22}
\end{figure}

In the short time domain subdiffusion is observed in the whole range of the orientational order parameter. Moreover, the slowdown of the diffusion process (as expressed by the value of the effective exponent $\alpha$) is stronger in more ordered phases, cf. Fig.\ref{fig:diffusion1}. This observation seems to be counterintuitive, as the motion in ideally ordered two dimensional space often reduces to a one dimensional problem (e.g. when formation of  channels favors unrestricted motion in a preferential direction) where standard diffusion should be observed. However, in the case analyzed here only nematic-like order is observed and additionally, ordering favors a denser fibrinogen packing \cite{bib:Ciesla2012c}, which results in more obstacles encountered by a diffusing molecule.
To analyze this phenomenon more carefully we treated diffusion as being composed of two separate GRW: parallel and perpendicular to the direction of the global orientational order:
\begin{equation}
 x^2 = x_{\perp}^2 + x_{\parallel}^2.
 \label{xperppar}
 \end{equation}
 Results obtained for those processes are presented in insets to Fig.\ref{fig:diffusion1}. The parallel diffusion speeds up with increasing order ($\alpha_{\parallel}>\alpha_{\perp}$), in contrast to the perpendicular  transport which becomes more obstructed and slows down. Both motions converge to the same exponent value  $\alpha_{\parallel}=\alpha_{\perp}$ at $q=0$, as in disordered phase there is no intrinsic preferred direction and the diffusion homogenizes. 
The effective exponent $\alpha$  depends on a tracer's size and the overall diffusive motion is slower for larger molecules,  as expected (see Figs.\ref{fig:diffusion1},\ref{fig:diffusion22}). For small tracers ($d\to 0$), parallel motion approaches normal diffusion  ($\alpha \to 1$) and this behaviour is stronger in ordered layers, see Fig.\ref{fig:diffusion22}. In contrast, low ordering implies higher degree of structural homogeneity of layers and results in almost equal values $\alpha_{\parallel}\approx\alpha_{\perp}$.

\subsection{Long time limit}
To get a deeper insight into the global diffusion process, longer trajectories of tracer movement have been analyzed. Additionally, structures of fibrinogen layers have been formed out of a fixed, constant concentration of molecules. Samples of surfaces created in this way are characterized by various spatial configurations of the same number of molecules distributed over the target area. In simulations  the maximum length of a single GRW trajectory has been preset to include $2\times 10^5$ steps and the 
MSD has been evaluated as an average over $10^3$ independent GRW  trajectories starting from randomly chosen, uncovered points.
\par
\begin{figure}[htb]
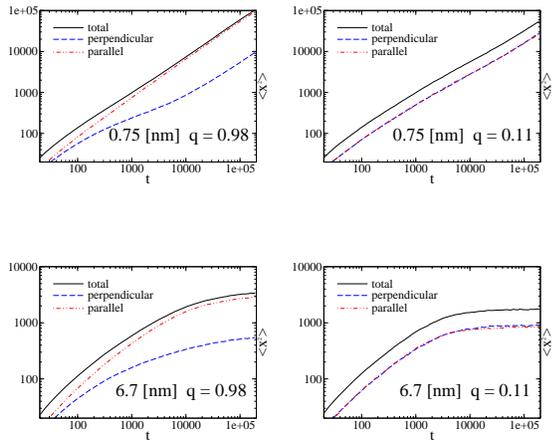

\vspace{1cm}
\centerline{%
\includegraphics[width=0.4\columnwidth]{diff_098}
\hspace{0.1cm}
\includegraphics[width=0.4\columnwidth]{diff_011}
}
\vspace{1cm}
\centerline{%
\includegraphics[width=0.4\columnwidth]{diff_098_large}
\hspace{0.1cm}
\includegraphics[width=0.4\columnwidth]{diff_011_large}
}
\caption{Tracer MSD dependence on time for ordered (left) and disordered (right) layer. The tracer molecules were spheres of diameter $0.75$ nm (top) and $6.7$ nm (bottom), respectively. Black lines indicate the total MSD, whereas red and green ones correspond to perpendicular and parallel diffusion, see Eq.(\ref{xperppar}). Simulations have been performed for a constant concentration of fibrinogen molecules on a sample surface, c=2140$\mu$m$^{-2}$. }
\label{fig:msd_time}
\end{figure}

The dependence of $\langle x^2 \rangle$ on number of steps for two different sizes of a spherical tracer is presented in Fig.\ref{fig:msd_time}.
Unlike for small tracer molecules, diffusive transport of large particles can be seen only in relatively short time intervals. The motion is then characterized by a sublinear MSD behavior (see Fig. \ref{fig:msd_time}) and gradually becomes effectively confined within the network of fibrinogen molecules adsorbed on the surface. At longer times ($t>10^4$) MSD saturates pointing to the immobilization of particles within the mesh. Consequently, in order to avoid the entrapment, further analysis was performed using only small tracers. Their GRW motion has been analyzed by inspecting  trajectories of duration $10^4<t<10^5$. Those longer trajectories exhibit sometimes the slowdown of the diffusion process related to a  finite size of the layers. 
Such cases,  whenever observed in simulations,  have been excluded from statistical analysis.
\par
\begin{figure}[htb]
\vspace{1cm}
\centerline{%
\includegraphics[width=0.4\columnwidth]{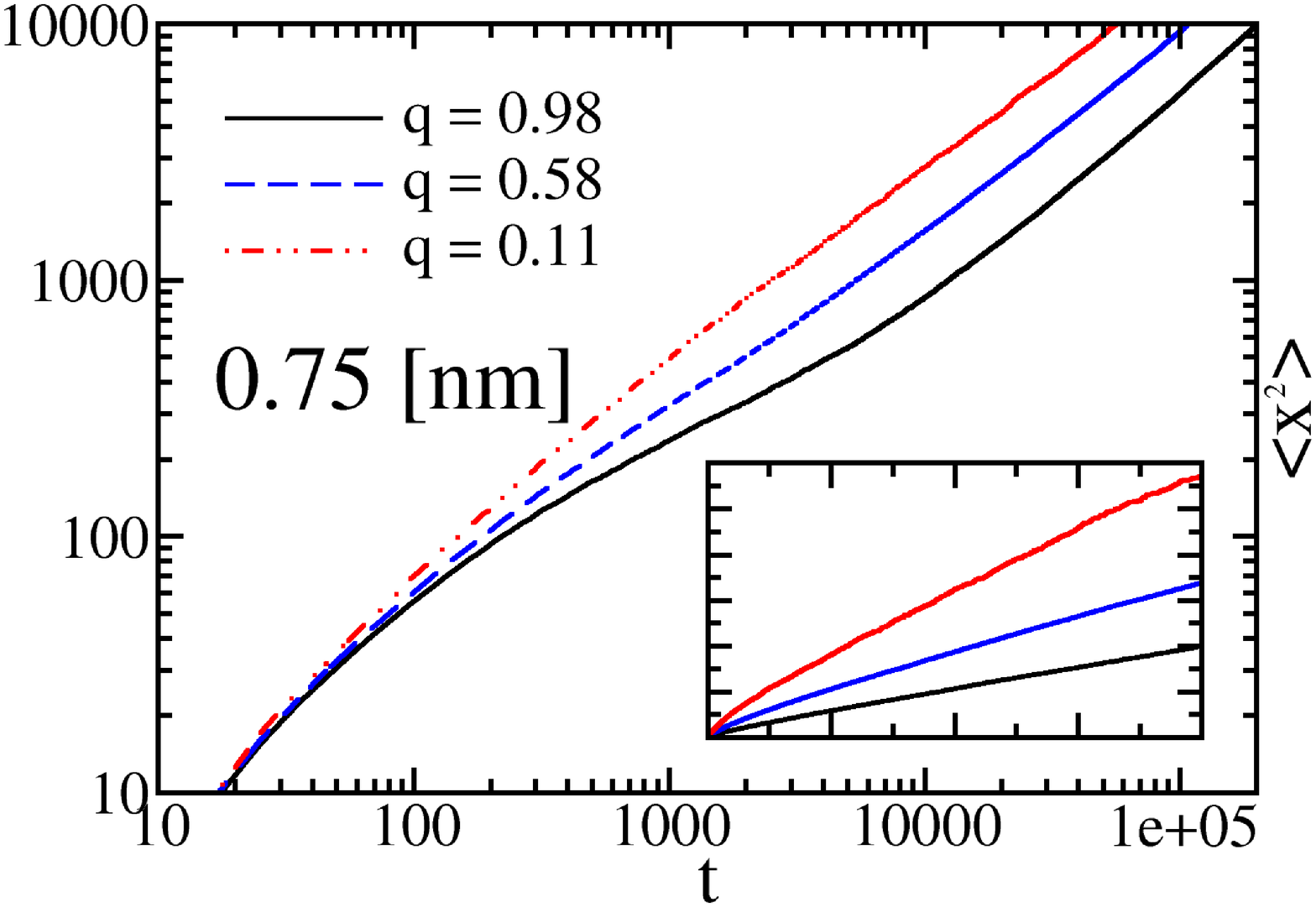}
\hspace{0.1cm}
\includegraphics[width=0.4\columnwidth]{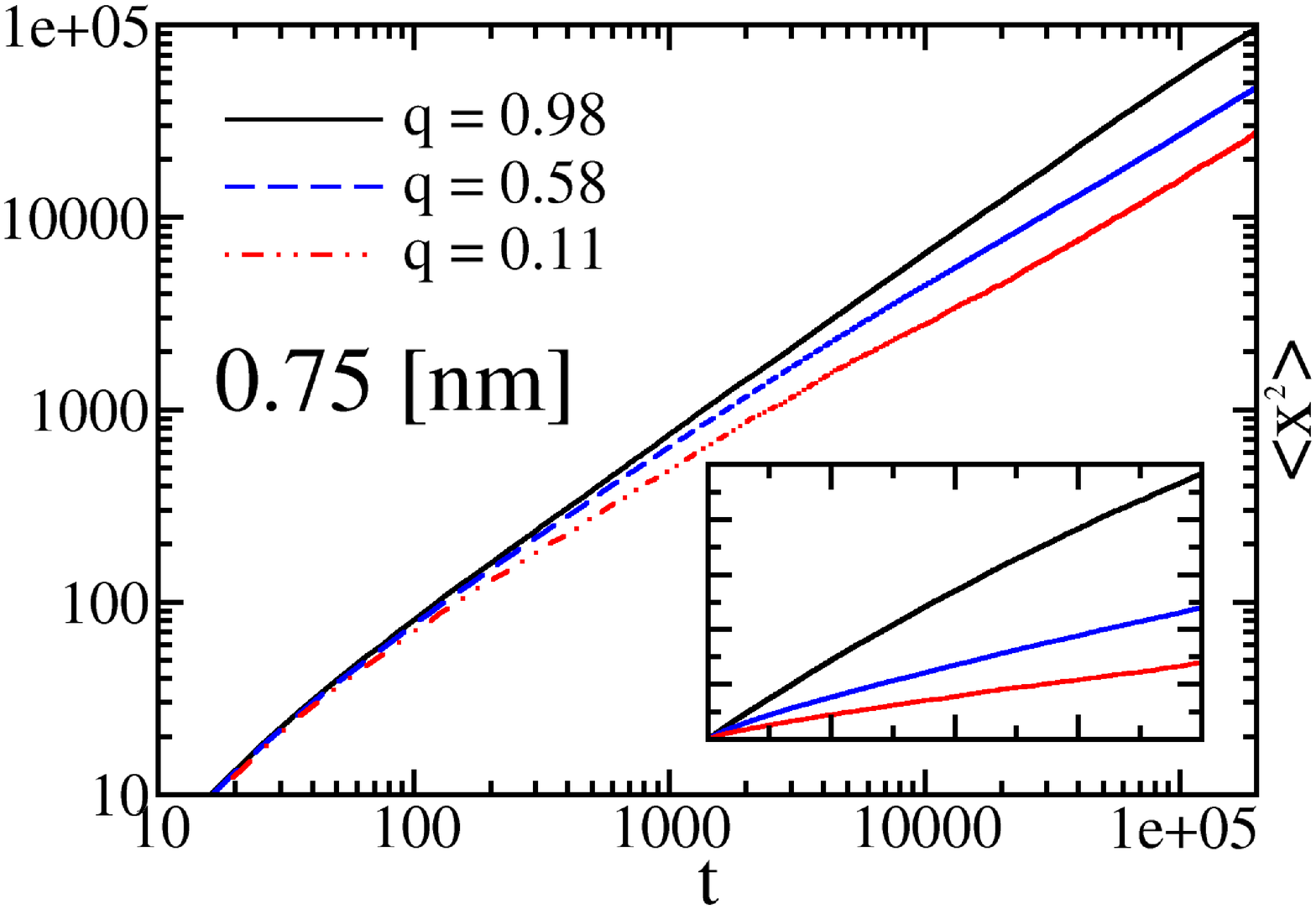}
}
\caption{Perpendicular (left) and parallel (right) components of the MSD for three different orientational order parameter $q$. Insets show the same data but in linear scale. The tracer diameter was $0.75$ nm and the concentration of fibrinogen molecules c= 2140$\mu$m$^{-2}$.}
\label{fig:perppar_time}
\end{figure}

The parallel and perpendicular MSD, defined by Eq.(\ref{xperppar}) are presented in Fig.\ref{fig:perppar_time}.
Inspection of Fig.\ref{fig:perppar_time} indicates that in ordered layers of fibrinogens the parallel component of MSD is promoted, whereas the perpendicular  motion becomes hindered, in line with observations drawn from the upper panel of Fig.\ref{fig:msd_time}.
\par
\begin{figure}[htb]
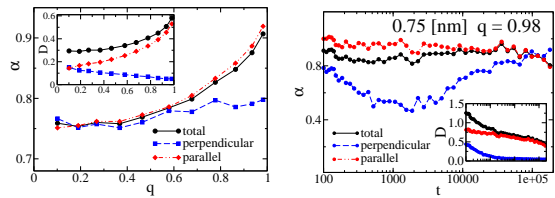

\vspace{1cm}
\centerline{%
\includegraphics[width=0.4\columnwidth]{diff_q}
\hspace{0.1cm}
\includegraphics[width=0.4\columnwidth]{D_A_t}
}
\caption{Dependence of  the diffusive transport on orientational order parameter and time. Left panel: exponent $\alpha$  derived from the fit to Eq.(2) as a function of orientational order $q$ inside a layer. The inset presents time-scale dependent diffusion coefficient $D(t)$ evaluated from Eq.(\ref{altD}). Parameters  $\alpha$, $D$ were averaged over $t \in [10^4, 10^5]$ time window. \\
Right panel: the temporary value (cf. Eq.(\ref{altA})) of the exponent $\alpha(t)$ is presented  for the order parameter $q=0.98$. The inset displays the time-scale dependent diffusion coefficient $D(t)$ evaluated for the same time window. In both cases the tracer diameter was $0.75$ nm and the concentration of obstacles $c=2140\; \mu\mbox{m}^{-2}$ .}
\label{fig:diffusion2}
\end{figure}

In order to further elucidate effects due to varying spatial anisotropy and geometry of the fibrinogen mesh, we focused our analysis on  the effective exponent $\alpha$ and coefficient $D$ in relation (\ref{diffusion}), (\ref{altD}) and (\ref{altA}). It's dependence on time and orientational order parameter is summarized in Fig.\ref{fig:diffusion2}.
Similarly to the case of transport observed in the short time domain (see Section \ref{short}),  subdiffusion is observed in the whole range of the orientational order parameter and the exponent $\alpha$ is highest for parallel MSD component in almost all range of $q's$. On the other hand, the diffusion coefficient is significantly lower than for the whole MSD, which reflects variability of $D$ and $\alpha$ parameters in perpendicular and parallel directions of motion: $\langle x^2 \rangle_{\perp} = D_{\perp} t^{\alpha_{\perp}}$ and $\langle x^2 \rangle_{\parallel} = D_{\parallel} t^{\alpha_{\parallel}}$.  
Moreover, for ordered phases ($q=0.98$) the  perpendicular exponent $\alpha_{\perp}$ has a broad minimum for intermediate times (see Fig.9, right panel) which corroborates 
 with observation of the plateau in the perpendicular component of MSD (see Fig.8, left panel). This minimum corresponds to the confinement of a tracer between long, parallel filaments made of fibrinogen molecules. Similar effect has been also detected in models of the anisotropic diffusion of a strongly hindered rod represented by a thin needle moving in an array of hard point obstacles \cite{bib:Munk2009}.
\section{Tracer diffusion as a percolation problem}

\begin{figure}[htb]
\centerline{%
\vspace{1cm}
\includegraphics[width=0.9\columnwidth]{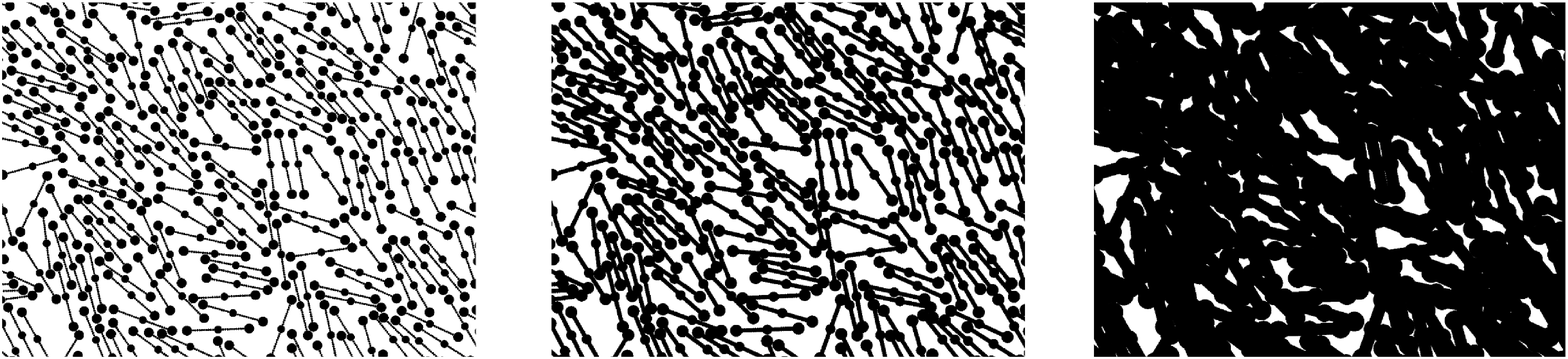}}
\caption{Obstacles seen from a perspective of the different sized tracer center. The tracer diameter is $0$ (left), $1.5$ nm (middle) and $6.7$ nm (right). The fibrinogen concentration is c=$2140\; \mu\mbox{m}^{-2}$ and $q=0.68$.}
\label{fig:effectivelayer}
\end{figure}

The model we have to do with is essentially a continuous percolation model \cite{KlaSo,Havlin}, but not a standard one. 
To see this, let us consider on the motion of a single tracer of radius $r = d/2$ and concentrate on the motion of its center. Since both the tracers and the fibrinogen molecules are modeled as solid, impenetrable objects, the motion of center of the tracer is geometrically restricted to the domain of the plane outside the $r$-vicinities of all fibrinogen molecules. Although the areas of the molecules themselves cannot intersect, their $r$-vicinities can, thus leading to creation of effectively impenetrable domains, see Fig.\ref{fig:effectivelayer}. 

\begin{figure}[htb]
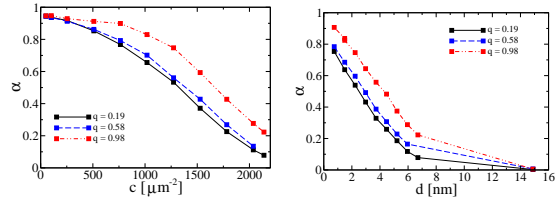

\hspace{1cm}
\centerline{%
\includegraphics[width=0.4\columnwidth]{alpha_q_conc}
\hspace{0.1cm}
\includegraphics[width=0.4\columnwidth]{alpha_q_d}}
\caption{Dependence of the effective diffusion exponent $\alpha$ on fibrinogen concentration (left) and tracer size (right) for three different orientational orders in a layer. The tracer diameter in the  left plot is $6.7$ nm, and the concentration in the right plot is $2140\, {\rm \mu}m^{-2}$. Evaluation of the exponent $\alpha$ has been performed on trajectories of duration of $10^4 < t < 10^5$.}
\label{fig:alpha_q_conc1}
\end{figure}

The percolation threshold is defined by a critical concentration of connected accessible points at which the unbounded motion through the lattice becomes possible. In our system the percolation threshold depends however in a complex manner on three control parameters: on the orientational parameter $q$,
on the concentration of fibrinogen molecules $c$ which itself depends on $q$ via our generation algorithm,
and on the tracer's diameter $d$.
Figs.\ref{fig:msd_time}, \ref{fig:alpha_q_conc1} show essentially the behavior of the MSD in the percolating and non-percolating cases. In the first case the MSDs continuously grow, in the second case they saturate at the value corresponding to the size of typical finite clusters (accessible areas). The transition here takes place due to both: increasing size of the tracer $d$ (Fig.\ref{fig:msd_time}) and increasing concentration of obstacles (Fig.\ref{fig:alpha_q_conc1}) for ordered and disordered situations. \\ \\

\subsection{Anisotropy}

\begin{figure}[htb]
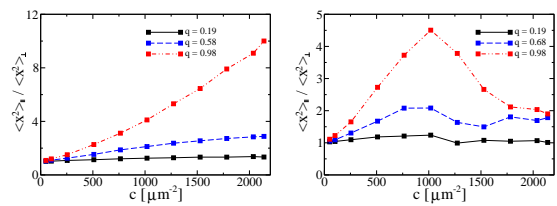

\vspace{1cm}
\centerline{%
\includegraphics[width=0.4\columnwidth]{parperp_cq_1}
\hspace{0.1cm}
\includegraphics[width=0.4\columnwidth]{parperp_cq_20}}
\caption{Dependence of the parallel to perpendicular MSD ratio (after $t=1.5 \times10^5$) on  fibrinogen concentration and orientational order in a layer. The tracer diameter is $0.75$ nm and $6.7$ nm for left/right panel, respectively.}
\label{fig:alpha_q_conc}
\end{figure}

In the percolating regime the system homogenizes, and the diffusion in it is normal, corresponding to the long-time value
of $\alpha = 1$, cf. upper panel of Fig.\ref{fig:msd_time} and Fig.\ref{fig:perppar_time}. However, this normal diffusion can be strongly anisotropic, as demonstrated in Figs.\ref{fig:msd_time}, \ref{fig:perppar_time}, \ref{fig:alpha_q_conc}.
The exact calculation of anisotropy is a hard task, and the results are known only either for small concentrations or for very simple
geometric forms of inclusions (ellipses or ellipsoids). Nevertheless, several general remarks can be made based on what is known for such simple cases \cite{KlaSo,Havlin}. 

If the obstacles are not oriented, no anisotropy can be observed. If they are oriented, the final diffusion coefficient is anisotropic. As a measure of anisotropy the 
quotient of the values of diffusivity (or, otherwise of the MSD components) parallel and perpendicular to the orientation of obstacles  \cite{bib:Hofling,Ba} can be taken: $A=D_{\parallel}/D_{\perp}$, which
changes in the interval $1 \leq A \leq \infty$.

The problem of finding the anisotropy of the diffusion coefficient in a system with obstacles is similar to the problem of finding the heat conductivity or the electric conductivity in similar settings. Such problems were investigated for several different
geometries of inclusions. For small concentration of obstacles the diffusivity $D$ of the system is well-described by the perturbative approaches \cite{Klemens,Ba}, and its anisotropy vanishes (which corresponds to $A=1$), as also seen in Fig.\ref{fig:alpha_q_conc}. 

For the case of intersecting obstacles with centers randomly placed on a plain the anisotropy close to the percolation threshold can grow  quite large. Thus, close to percolation transition the anisotropy in a system of highly oriented elongated defects might be approximated by those of a system with ellipsoidal inclusions of the same aspect ratio $a/b$ with $a$ being the largest and $b$ the smallest dimension of the inclusion, a problem discussed in \cite{Shklovskii}.
The discussion in this work is essentially for 3d, but can be easily adapted to a two-dimensional case. 
In this case close to the criticality $A \approx(a/b)^2$, and by taking geometrical parameters of the fibrinogen model, may get to be as large as $10^2$ in our case. Note that the possibility of intersection and the random position 
of the centers of inclusions are both crucial for the discussion, and the departures from these assumptions may lead to strong deviations from this prediction in what concerns the anisotropy's concentration dependence. 
Thus, in the case of parallel non-intersecting inclusions  with large aspect ratio ("scratches'') of length $a$ thrown onto a plane with concentration  $c$ one gets $A = (\pi a^2 c)^2$, see \cite{Ba2}. Our situation here is more complex than the both ones considered theoretically due to the fact that the domains are
not fully oriented, can intersect but possess hard cores, so that their centers are not randomly distributed. This leads to highly 
non-monotonous concentration dependencies for the highly oriented case, as seen in the lower panel of Fig.\ref{fig:alpha_q_conc}. 

\subsection{Effective exponent of anomalous diffusion}

The overall accessible volume of the mesh close to the percolation threshold forms a fractal-like percolation cluster. The effective exponent  $\alpha$ is connected with a critical exponent of conductivity in percolation theory. The last one is either determined by the lattice percolation exponent $\mu$ or by the distribution of the resistances of the bottleneck (narrow paths between the almost contacting obstacles) \cite{Feng}. Despite much more complicated geometry of our model, the bottlenecks in it appear close to the contact points of essentially circular areas, and therefore should show the same resistivity singularity behavior as the Swiss cheese model \cite{Feng}
which has the same conductivity exponent as the lattice percolation. This one is translated into the standard value of the anomalous effective diffusion exponent being $\alpha^* \approx 0.70$ as defined through the known values of
spectral and fractal dimensions of the percolation clusters \cite{KlaSo,Novak}.

This value of the final exponent of anomalous diffusion has to be the same for both principal values of the displacement (i.e. in the directions parallel and perpendicular to the obstacles' orientation), and is observed to happen for orientational order $q = 0.58$ in our case (cf. Fig.\ref{fig:perppar_time},\ref{fig:diffusion2}). The behavior of each of the components for the values of parameters corresponding to the percolating phase is also quite typical: The normal short-time diffusion with the seed diffusivity coefficient (corresponding to practically non-constrained particle motion), a crossover to a more or less pronounced intermediate asymptotic subdiffusive behavior typical for percolation, and another crossover back to normal diffusion with the effective diffusion coefficient considerably smaller than the seed one (cf. left panels of Figs.7,8). In the non-percolating domain the last crossover is the one to a saturating MSD. 

However, Fig.\ref{fig:perppar_time} also shows that the positions of the crossover regions 
and the widths of the crossover domains for the two components are quite different. 
For example, for $q=0.982$ the perpendicular component already underwent 
both crossovers at $t=10^5$, while the parallel component has just entered the first crossover domain at the same time. Summing up two functions with so strongly different 
behaviors to obtain the total diffusivity leads to the effective behavior which is seemingly far from typical percolation predictions. Thus, in the  Figs.\ref{fig:diffusion1}, \ref{fig:diffusion22} the exponents $\alpha$ of single components of MSD are either larger than or equal to (within the error bars) the percolation value of $\alpha^*$, as it has to be the case for the typical double crossover situation. The exponent characterizing the total displacement  can however
 fall below the value of 0.55, thus absolutely obscuring the percolative nature of the behavior observed.  

\section{Conclusions}

In this paper properties  of diffusive motion of a tracer particle in crowded, anisotropic environment has been analyzed. In general, transport slowdown  with the characteristic effective exponent $\alpha<1$ has been observed. However, details of the process depend in a complex way on three control parameters: concentration of obstacles, their orientational (nematic-like) ordering, as well as on the size of a tracer. Moreover, due to environment anisotropy, the perpendicular and parallel motion to the intrinsic direction represented by the director orientation, cf. Eq.(\ref{order}) has to be studied separately, as both components of the motion can exhibit a totally different behaviour.

 Although our situation is definitely a variant of a continuous percolation model, it shows a very rich and interesting behavior due to its novel aspect, namely the anisotropy. The characteristic crossover times for the components of motion parallel and perpendicular to the main axes can be very different, so that the total diffusion coefficient can show effective time dependence which does not resemble the one for lattice percolation, and can even exhibit much slower effective subdiffusive behavior on intermediate times.
 
 Analysis of the effective exponent $\alpha$ averaged over the time window  $t \in [10^4, 10^5]$ shows that subdiffusive transport of small tracers is observed in the whole range of the orientational order parameter and the expected crossover to percolation limit is detected for $q\approx 0.58$, see Fig.9. For high orientational order, closer inspection of  time dependent $\alpha(t)$ allows to detect another crossover $\alpha_{\perp}=\alpha_{\parallel}\approx 0.9$ for very long times $t\approx10^5$. Yet, this "almost normal" diffusion remains strongly anisotropic, ($D_{\parallel}\neq D_{\perp}$) as demonstrated in the right panel of Fig.9 and left panel of Fig.12. In general, positions of crossover regions and duration of the crossover domains are different for two orientational components of motion and can be rationalized in terms of transport on an anisotropic percolating space.
 
We expect our findings to be relevant in understanding sources of anomalous transport in crowded, heterogeneous environments. Active research in this field is beneficial for control of anomalous subdiffusion in biological media, calibration of measurements by use of fluorescence correlation spectroscopy (FCS) and analysis of single particle tracking within the dense environment of the cell \cite{bib:Hofling,Garini,Holyst}. It is expected also to contribute to further progress in a general theory \cite{Majka1,Majka2,bib:Wang} of the anisotropic motion of objects in entangled suspensions responsible for steric effects.
\section*{Acknowledgments}
This work was supported in part by European Science Foundation EPSD grant. Francesc Sagu\'es acknowledges financial support from MICINN (FIS2010-21924-C02-01) and
from DURSI Generalitat de Catalunya (2009SGR 1055). The authors acknowledge many fruitful discussions with Jakub Barbasz.
%

\end{document}